# Generation of Vector Platicons and Hybrid Soliton-Platicon Complexes in Optical Microresonators via Modulated Pump

Valery E. Lobanov[1], Artem E. Shitikov[1], Ramzil R. Galiev[1,2], Kirill N. Min'kov[1], Olga V. Borovkova[2], Nikita M. Kondratiev[1]

[1]*Russian Quantum Center, Skolkovo, 143026, Russia*
[2]*Faculty of Physics, Lomonosov Moscow State University, Moscow 119991, Russia*
v.lobanov@rqc.ru

**Abstract:** We demonstrate numerically the possibility of generation of vector platicons having orthogonally polarized components by the modulated pump at normal GVD. Dynamics of this process can be controlled by pump polarization and depends on both the spectral interval between pumped modes and the difference of their FSRs. We also show the possibility of the excitation of the hybrid soliton-platicon complexes if pumped modes have alternating signs of the GVD coefficient. © 2021 The Author(s)

**I. INTRODUCTION**

To date, generation and properties of different types of solitonic pulses, including dissipative Kerr solitons [1] at anomalous group velocity dispersion (GVD) and dark pulses or platicons [2, 3] at normal GVD, are well-studied in high-Q optical microresonators, both bulk and on-chip [4-8]. Such coherent optical frequency combs are actively used in different areas of science and technology such as high-precision metrology [9, 10], high-resolution spectroscopy [11,12], astrophysics [13,14], and high-volume telecommunication systems [15-17]. Recently, more complex nonlinear localized structures, for example, multicomponent ones, have attracted the attention of researchers. It has been revealed that nonlinear interaction between the fields having different transverse profiles or different polarizations provides interesting opportunities for generation of frequency combs and solitons [18-21]. In particular, a generation of complex solitonic structures existing due to the nonlinear coupling between orthogonally polarized fields [22-25] has been shown. Such approach has been studied for bright solitons at anomalous GVD [22-24] and platicons or dark solitons at normal GVD [25,26]. Since generation of conventional platicons is significantly more efficient than the generation of bright solitons in terms of the pump-to-comb conversion efficiency [27,28], the generation of multicomponent platicons seems to be very interesting for many actual applications, e.g. coherent optical communications. In [25] generation of such solitonic structures was demonstrated numerically by means of two pump beams which can be bulky in experiment and hard to realize in some cases. Here we demonstrate numerically that under certain conditions one may use single amplitude-modulated pump beam with the frequency of the amplitude modulation equal to one free spectral range (FSR) of the microresonator [29, 30] for the generation of vector platicons with orthogonally polarized components. Dynamics of this process can be controlled by pump polarization and depends on the distance between orthogonally polarized modes and on the difference of their FSRs. We also show that in more exotic case, when one mode family has anomalous GVD, but another experiences normal GVD, excitation of nonlinearly coupled soliton-platicon complexes is also possible.

## II. MODEL

We used a model, based on the system of two coupled LLE-like equations [20], that takes into account a nonlinear cross-action of the orthogonally polarized fields and amplitude modulation of pump $f(t) = f(1 + \varepsilon \cos \Omega t)$ [29, 30]:

$$\begin{cases} \dfrac{\partial \psi_1}{\partial \tau} + \delta_1 \dfrac{\partial \psi_1}{\partial \varphi} = \dfrac{id_{21}}{2}\dfrac{\partial^2 \psi_1}{\partial \varphi^2} + i\psi_1\left(|\psi_1|^2 + \dfrac{2}{3}|\psi_2|^2\right) - (1+i\alpha)\psi_1 + f(\varepsilon,\varphi)\cos\theta, \\ \dfrac{\partial \psi_2}{\partial \tau} + \delta_2 \dfrac{\partial \psi_2}{\partial \varphi} = \dfrac{id_{22}}{2}\dfrac{\partial^2 \psi_2}{\partial \varphi^2} + i\psi_2\left(|\psi_2|^2 + \dfrac{2}{3}|\psi_1|^2\right) - \left(\dfrac{\kappa_2}{\kappa_1}+i(\alpha-\Delta)\right)\psi_2 + f(\varepsilon,\varphi)\sin\theta, \end{cases} \quad (1)$$

where $\psi_{1,2}$ stand for the normalized waveforms of the corresponding orthogonally polarized waves, $\tau = \kappa_1 t/2$ is the time, normalized to the first mode photon lifetime, $\kappa_{1,2}$ are the cavity decay rates for the first and second pumped modes (or the loaded linewidths of the considered microresonator modes), $\varphi \in [-\pi;\pi]$ is the azimuthal angle in a coordinate system rotating with the angular frequency equal to the pump modulation frequency $\Omega$. For the considered microresonator mode families the microresonator eigenfrequencies are assumed to be $\omega_{\mu_1} = \omega_{01} + D_{11}\mu + \dfrac{1}{2}D_{21}\mu^2$ and $\omega_{\mu_2} = \omega_{02} + D_{12}\mu + \dfrac{1}{2}D_{22}\mu^2$, where $\omega_{01}$ and $\omega_{02}$ are the eigenfrequencies of the first and second pumped mode, respectively. All mode numbers $\mu_{1,2}$ are defined relative to the pumped modes with $\mu_{1,2} = 0$. $D_{11,12}$ and $D_{21,22}$ are free spectral ranges and GVD coefficients for the first and second mode families, respectively; $d_{21,22} = 2D_{21,22}/\kappa_1$ are the normalized GVD coefficients (positive/negative GVD coefficients correspond here to the anomalous/normal GVD, respectively); $\Delta = 2(\omega_{01} - \omega_{02})/\kappa_1$ is the normalized spectral interval between orthogonally polarized pumped modes; $\alpha = 2(\omega_{01} - \omega_{pump})/\kappa_1$ is the normalized detuning of the pump frequency $\omega_p$ from the first mode eigenfrequency $\omega_{01}$. $f(\varepsilon,\varphi) = f(1+\varepsilon\cos\varphi)$ for 1-FSR amplitude modulation ($\Omega \approx D_{11}$) [29, 30], $\varepsilon$ is the modulation depth,

$f = \sqrt{\dfrac{8g\eta P_0}{\kappa_1^2 \hbar \omega_{01}}}$ is the dimensionless pump amplitude, $g = \dfrac{\hbar \omega_{01}^2 c n_2}{n_0^2 V_{eff}}$ is the nonlinear coupling coefficient, $V_{eff}$ is the effective mode volume, $n_2$ is the nonlinear refractive index, $\eta$ is the coupling efficiency ($\eta = 1/2$ for the critical coupling). $\delta_{1,2} = 2(D_{11,12} - \Omega)/\kappa_1$ are the normalized offsets of the modulation frequency $\Omega$ from the FSRs $D_{11,12}$ of the corresponding mode, $\theta$ is the polarization angle of the pump beam defined relatively to the polarization directions of the pumped modes ($\theta = 0$ corresponds to the pumping of the first mode only, $\theta = \pi/2$ – to the second mode pumping). Varying $\theta$ one may control the effective pump of the first and second modes. Note the 2/3 coefficient in the cross-action term due to the orthogonal polarization of the interacting fields. We also considered the linear coupling effects and the corresponding modifications of the dispersion laws to be negligible.

### III. FREQUENCY SCAN APPROACH

First, using Eqs. 1 we studied the process of platicon generation from a noise-like input at $\kappa_2/\kappa_1 = 1$, $\varepsilon = 0.4$ and normal GVD ($d_{21,22} = -0.02$) by means of the conventional frequency scan approach (see Fig. 1). This method is based on the pump frequency scan across the microresonator resonance [$\omega_{pump}(t) = \omega_{pump}(0) - \Upsilon t$, $\Upsilon$ is the scan speed] and is widely used in experiments for the dissipative Kerr solitons generation [1, 5]. Also, it was shown that such approach can be used for the generation of platicons by the modulated pump [29-31]. In numerical simulations we introduced linear-in-time variation of the pump frequency detuning $\alpha$: $\alpha(\tau) = \alpha(0) + v\tau$, where $v = 4\Upsilon/\kappa_1^2$ is the normalized pump frequency scan rate.

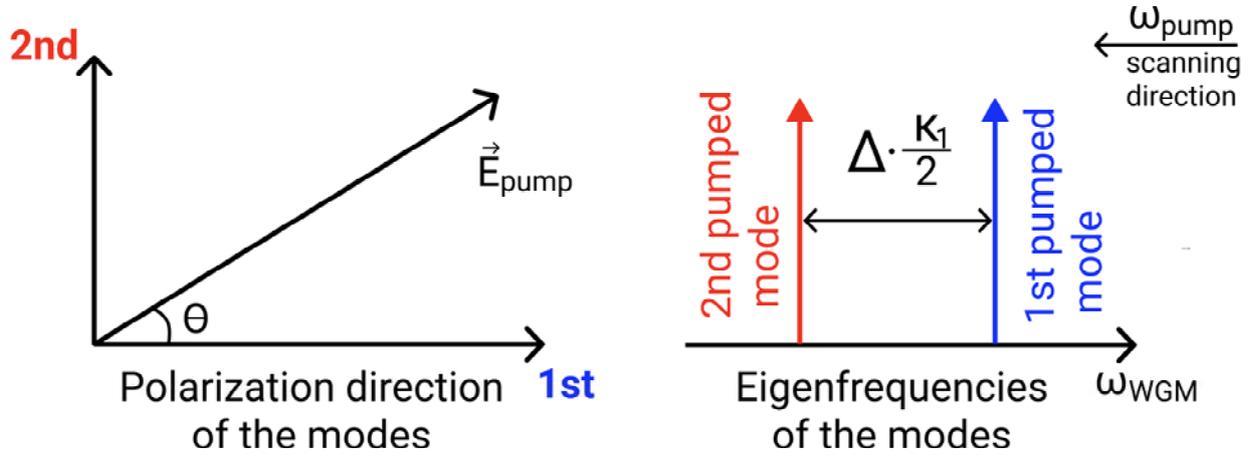

**Fig. 1.** Scheme of the process of the vector platicon generation by the simultaneous pump of the spectrally separated orthogonally polarized modes by the amplitude-modulated pump beam with linear-in-time decrease of the pump frequency. $\theta$ is the polarization angle of the pump beam, defined relatively to the polarization directions of the pumped modes. $\Delta$ is the spectral interval between orthogonally polarized pumped modes, normalized to $\kappa_1/2$.

First, we considered the case of $\theta = 0$, when only the first mode is pumped, and $\delta_1 = 0$ that means that modulation frequency is equal to the FSR of the first pumped mode. We used a rather low scan rate ($v = 0.0025$) and it was checked that dynamics of the considered processes does not change if $v$ decreases further. In the left panel in Fig. 2 one may see sudden power drop at $\alpha \approx 14.0$ corresponding to the abrupt change of the field distribution and formation of the localized state clearly visible in right panel. This localized state is a flat-top solitonic pulse or platicon. It exists at particular range of the pump frequency detuning $\alpha$ defined by the pump power and modulation depth. Platicons can be also considered as bound states of opposing switching waves [32] that connect upper and lower branches of bistable resonance to satisfy periodic boundary conditions. The width of the platicon gradually decreases with increasing pump frequency detuning. Generation of such structures is significantly more efficient than the generation of bright solitons in terms of the pump-to-comb conversion efficiency [27,28]; however, due to the absence of the modulational instability it requires special approaches besides the pump frequency scan, such pump modulation described here or controllable mode interactions [28, 33, 34] or self-injection locking [35, 36].

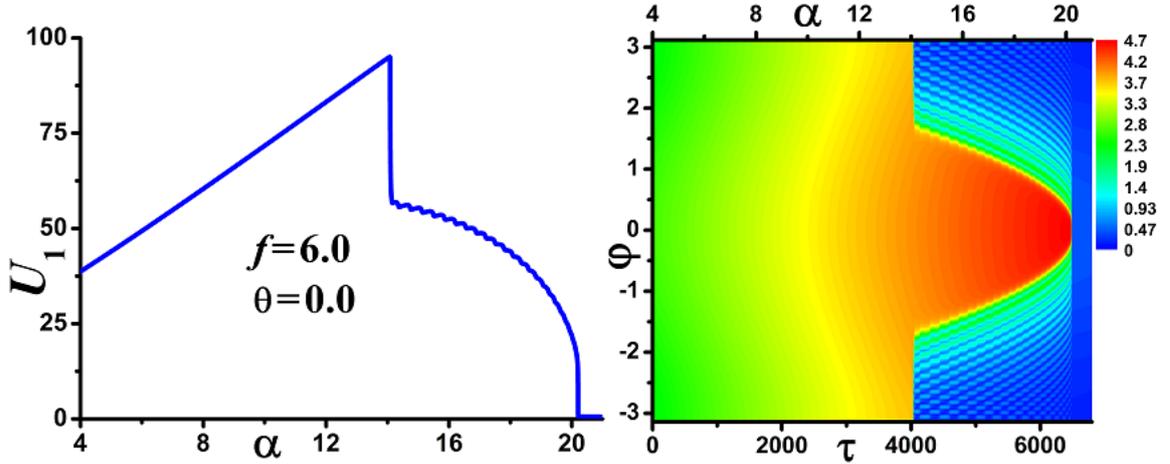

**Fig. 2.** Evolution of the intracavity power (left panel) and field distribution (right panel) upon pump frequency scan at $\theta = 0.0$, $d_{21} = -0.02$, $f = 6.0$, $\varepsilon = 0.4$, $\delta_1 = 0.0$, $v = 0.0025$. All quantities are plotted in dimensionless units.

If $0 < \theta < \pi/2$ then both modes are pumped simultaneously and the dynamics of the nonlinear processes arising upon pump frequency scan is more complicated. To analyze it one may study the evolution of the intracavity power or field distribution evolution for both polarization states upon pump frequency since abrupt changes of the amplitude profiles of the generated signals manifest themselves in pronounced changes of the intracavity power (see Fig. 2).

## IV. VECTOR PLATICONS

For different combinations of the polarization angle $\theta$ and interval $\Delta$ we revealed several possible regimes arising upon pump frequency scan: 1) simultaneous generation and decay of nonlinearly coupled platicons at both polarizations (full trapping); by analogy with vector solitons [37], further we will call such solitonic structures having two orthogonally polarized platicon components as vector platicon; 2) overlapping platicon generation regions; 3) independent generation of platicons at different polarizations. This is clearly seen in Fig. 3, illustrating evolution of the intracavity power for both polarization states upon pump frequency scan for different values of the spectral interval between pumped mode $\Delta$. The evolution of the intracavity power becomes more complicated in comparison with the case of the single pumped mode shown in Fig. 2.

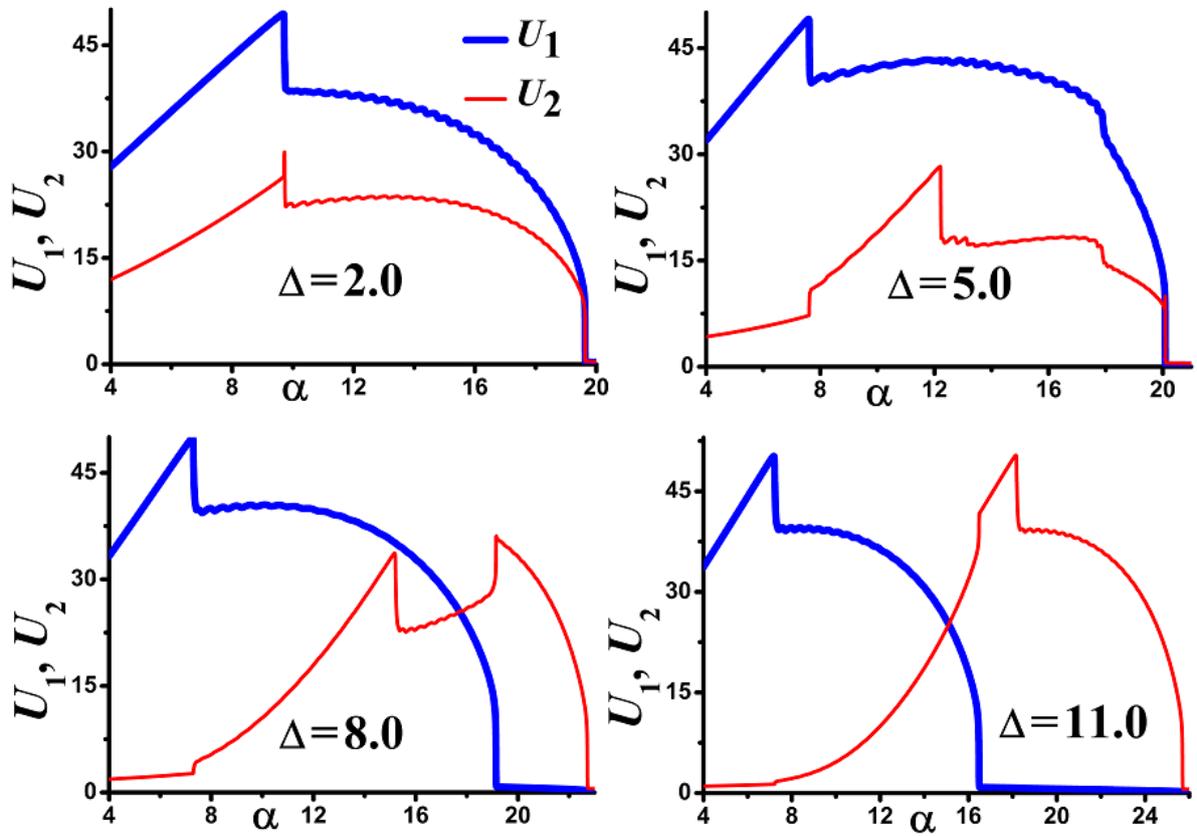

**Fig. 3.** Evolution of the intracavity powers for both polarization states upon pump frequency scan for different values of the interval between pumped modes $\Delta$ at $\theta = 0.25\pi$, $d_{21} = d_{22} = -0.02$, $f = 6.0$, $\varepsilon = 0.4$ $\nu = 0.0025$, $\delta_{1,2} = 0$. All quantities are plotted in dimensionless units.

One can observe the transformation of full trapping (top left panel, first drop of the upper thick blue line corresponds to the platicon generation) into the partial trapping (top right panel and bottom left panel, first drop of the lower thin red line occurs after the first drop of the thick blue line but before its second drop, corresponding to platicon decay) and then into independent platicon generation (bottom right panel) with the growth of the interval $\Delta$.

Field distribution and spectrum evolution for both polarization states for the same parameters as at top right panel in Fig. 3 are shown in Fig. 4. One may notice that abrupt changes of the intracavity power observed in Fig. 3 (see top right panel) correspond to the modifications of the field distribution. First power drop at $\alpha \approx 8$ corresponds to the excitation of the first-mode platicon; second power drop at $\alpha \approx 12$ – to the excitation of the second-mode platicon; weak drop at $\alpha \approx 18$ – to the rapid narrowing of the platicons (compare top right panel in Fig. 4 and right panel in Fig. 2). Interestingly, similar abrupt jumps of platicon parameters were observed in microresonators with backscattering due to the nonlinear interaction of the forward and backward waves [31]. Studying spectrum evolution (see bottom panels in Fig. 4) one can see that platicon generation at each component due to the nonlinear coupling leads to the spectrum broadening for both components and the widest spectrum is observed in the vector platicon state.

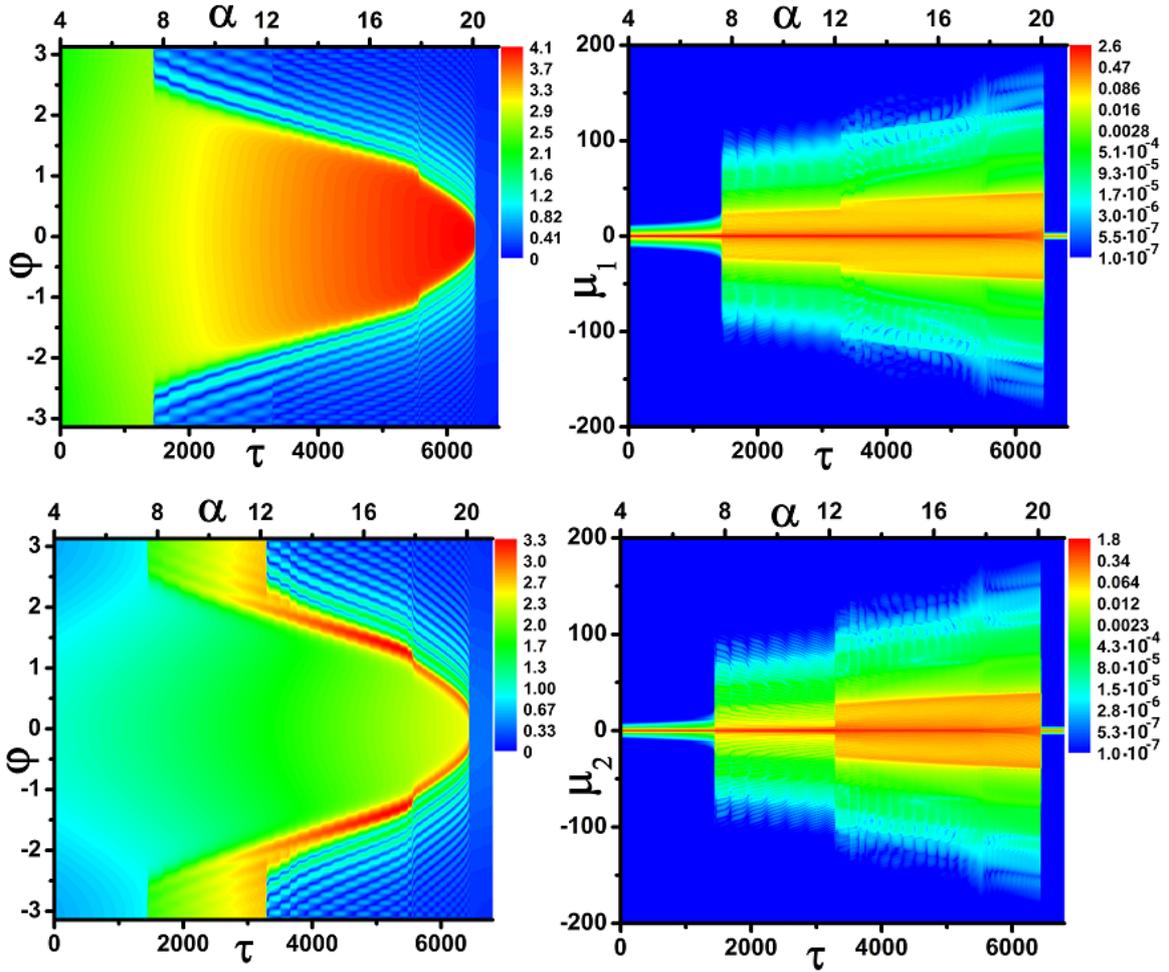

**Fig. 4.** (upper panels) First component and (bottom panels) second component field distribution (left column) and spectrum (right column, logarithmic scale) evolution upon forward pump frequency scan at $\theta = 0.25\pi$, $d_{21} = d_{22} = -0.02$, $f = 6.0$ $\varepsilon = 0.4$, $v = 0.0025$, $\Delta = 5.0$ $\delta_{1,2} = 0$ (see top right panel in Fig. 3). $\mu_{1,2}$ are mode numbers, $\mu_{1,2} = 0$ correspond to the pumped modes. All quantities are plotted in dimensionless units.

It was also found that nonlinear coupling leads to the distortion of the profiles of the generated signals. It is possible to identify several types of signals generated upon frequency scan (see Fig. 5 for the generated profiles for the process at the top right panel of Fig. 3 and in Fig. 4): i) high-intensity or low-intensity low-contrast signals at both polarizations (top left panel in Fig. 5); ii) platicon at one polarization and high-intensity or low-intensity low-contrast signal at another (top right panel); iii) nonlinearly coupled platicons at both polarizations (bottom panels). We also checked that in the absence of the frequency scan generated patterns propagate in a stable fashion over indefinitely large periods of time.

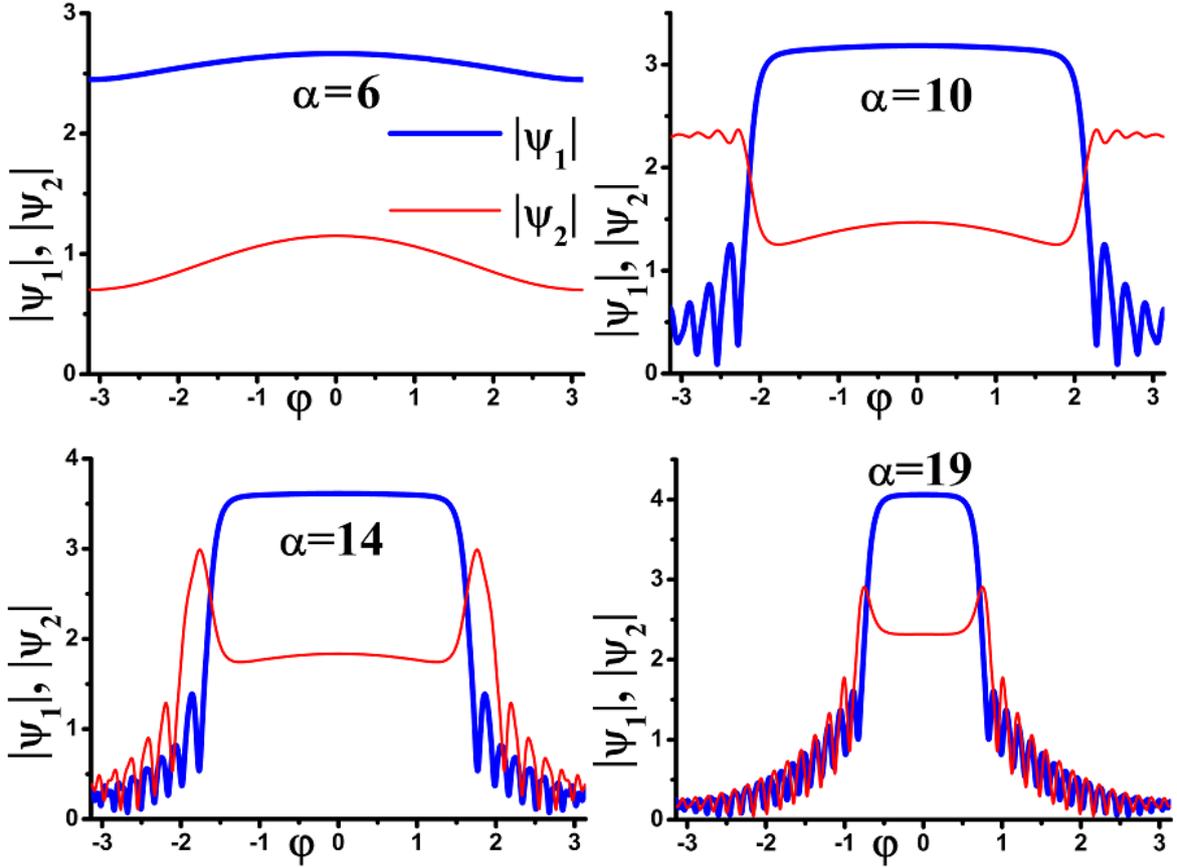

**Fig. 5**. Examples of the signals generated upon pump frequency scan for different values of $\alpha$ at $\theta = 0.25\pi$, $d_{21} = d_{22} = -0.02$, $f = 6.0$ $\varepsilon = 0.4$, $v = 0.0025$, $\Delta = 5.0$, $\delta_{1,2} = 0$. Field distribution and spectrum evolution for this process are shown in Fig. 4. All quantities are plotted in dimensionless units.

One may notice that platicon generated at the first pumped mode induce a dip at low-contrast profile of the signal at another polarization (top right panel in Fig. 5). The width of the dip is equal to the width of the platicon. Also, one may observe the same dip at platicon profile of another component appearing due to the nonlinear coupling of vector platicon components (see bottom panels in Fig. 5). Interestingly, nonlinear coupling is more pronounced at high-intensity parts of the soliton profiles, but it is rather weak at oscillating tails. It is clear seen at bottom panels in Fig. 5 that platicon tails for different components oscillate at different frequencies defined by the effective pump frequency detuning for each component.

The width and depth of the dip depend on the interval $\Delta$ (see Fig. 6). Note, that the dip usually appears at the wider platicon component, not at the less powerful one (compare top left and bottom left panels in Fig. 6). If orthogonally polarized modes have different thermorefractive coefficients, one may tune the value of $\Delta$ by varying microresonator temperature [38-40].

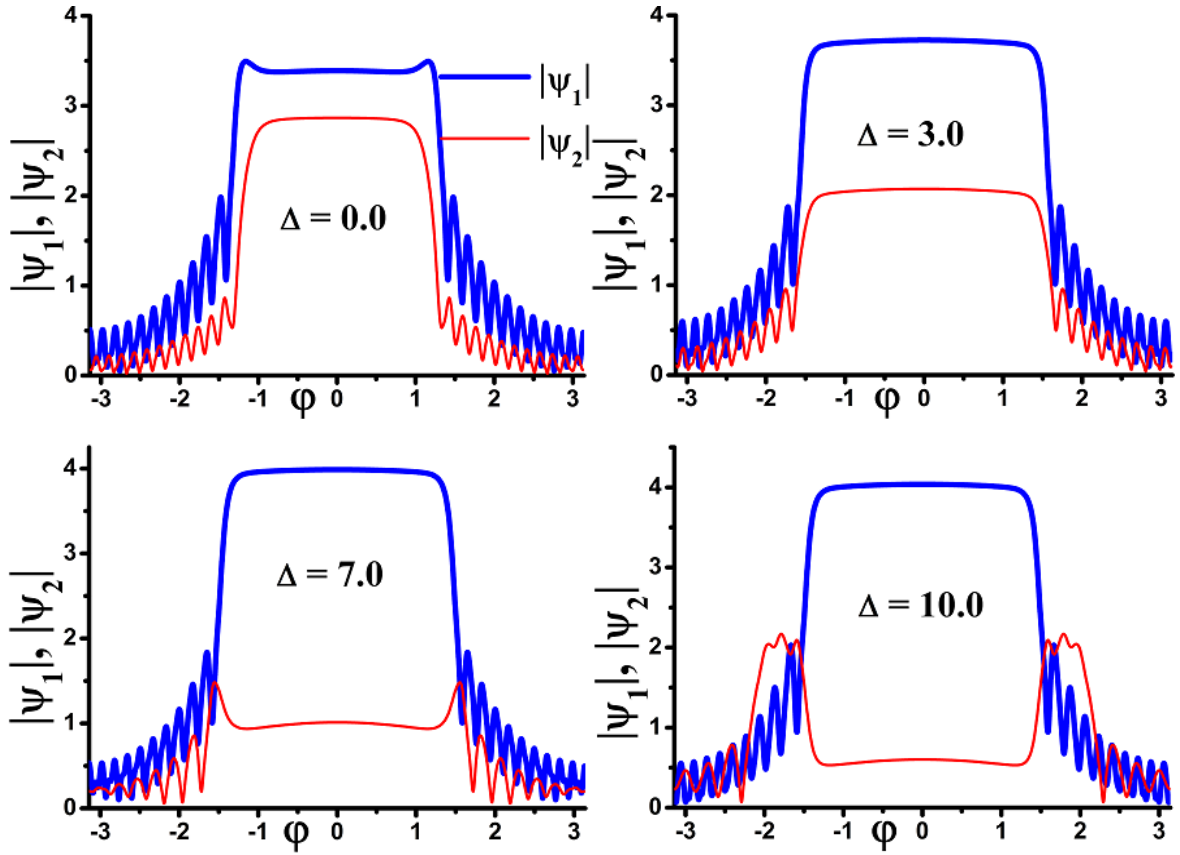

**Fig. 6.** Vector platicon components profiles for different value of $\Delta$ at $\theta = 0.15\pi$, $d_{21} = d_{22} = -0.02$, $f = 6.0$ $\varepsilon = 0.4$, $\delta_{1,2} = 0$, $\alpha = 15$. All quantities are plotted in dimensionless units.

The same transition from the simultaneous platicon generation to the independent platicon generation takes place with the growth of the polarization angle θ (see Fig. 7).

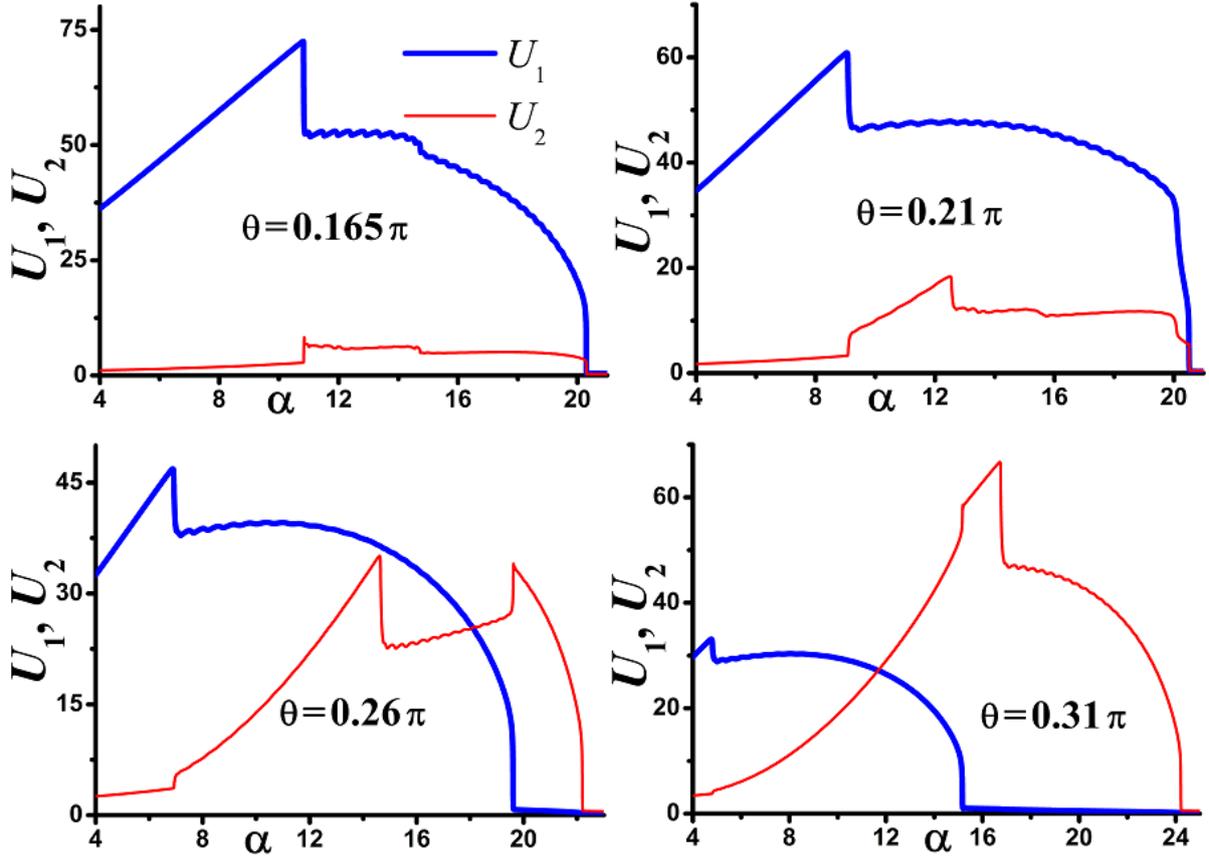

**Fig. 7.** Evolution of the intracavity powers for both polarization states upon pump frequency scan for different values of the polarization angle $\theta$ at $\Delta = 7.0$, $d_{21} = d_{22} = -0.02$, $f = 6.0$ $\varepsilon = 0.4$, $v = 0.0025$, $\delta_{1,2} = 0$. All quantities are plotted in dimensionless units.

If vector platicon is generated at one pump polarization angle $\theta$, in some cases one can control its parameters slowly varying it (see Fig. 8). Thus, pump polarization provides additional degree of freedom for the efficient control of properties of the generated coherent frequency comb.

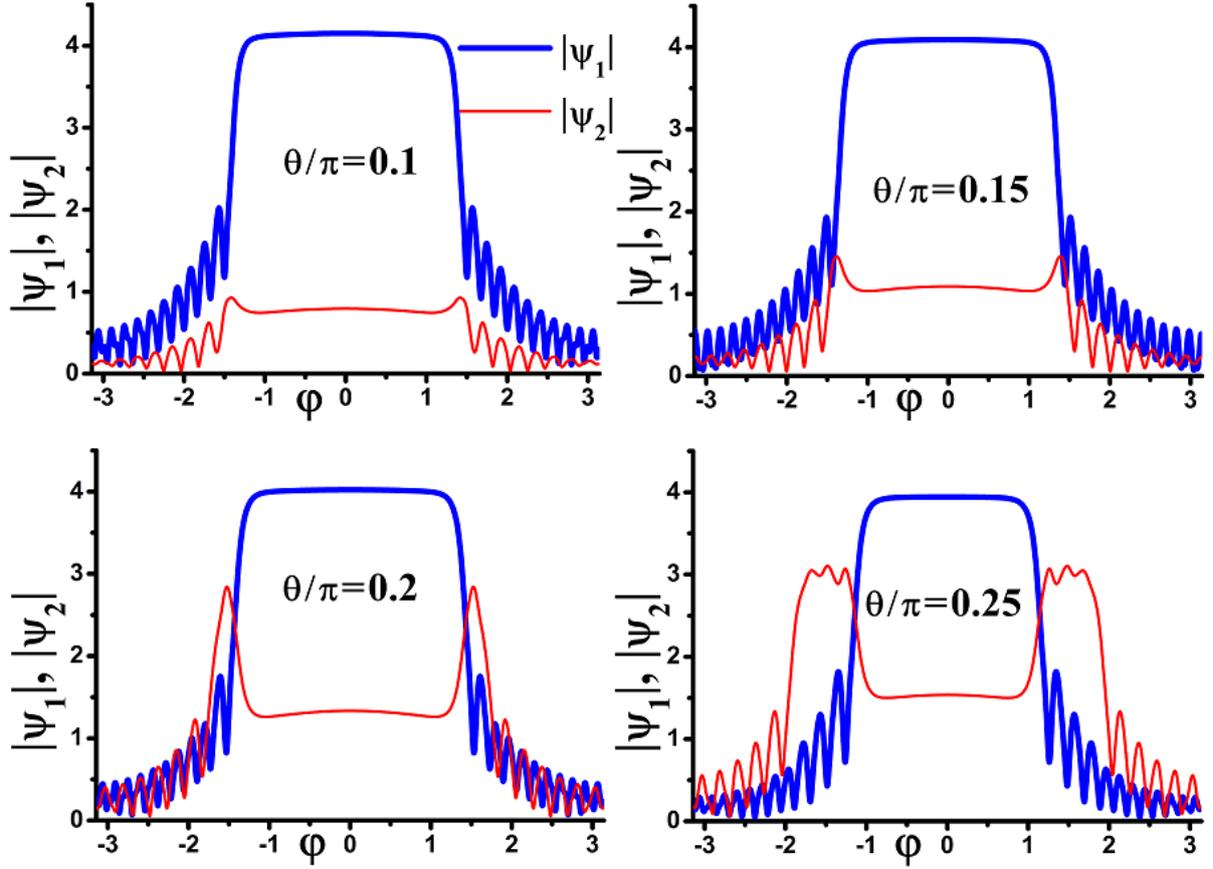

**Fig. 8.** Vector platicon components profiles for different value of $\theta$ at $\Delta = 7.0$, $d_{21} = d_{22} = -0.02$, $f = 6.0$ $\varepsilon = 0.4$, $\delta_{1,2} = 0$, $\alpha = 16$. All quantities are plotted in dimensionless units.

Then, we studied in more detail the parameter range providing generation of the vector platicons. It was revealed that the spectral and polarization angle ranges, where generation of the vector solitons is possible, become narrower with the growth of the interval between pumped modes (see Fig. 9). The critical value of the interval between pumped modes for the vector platicon generation was found to decrease monotonically with the growth of the pump polarization angle (see bottom right panel in Fig. 9). For considered parameters the generation of the vector platicons was observed for $\Delta < 20$, that means that frequency interval between pumped modes should be less than $10\kappa_1$.

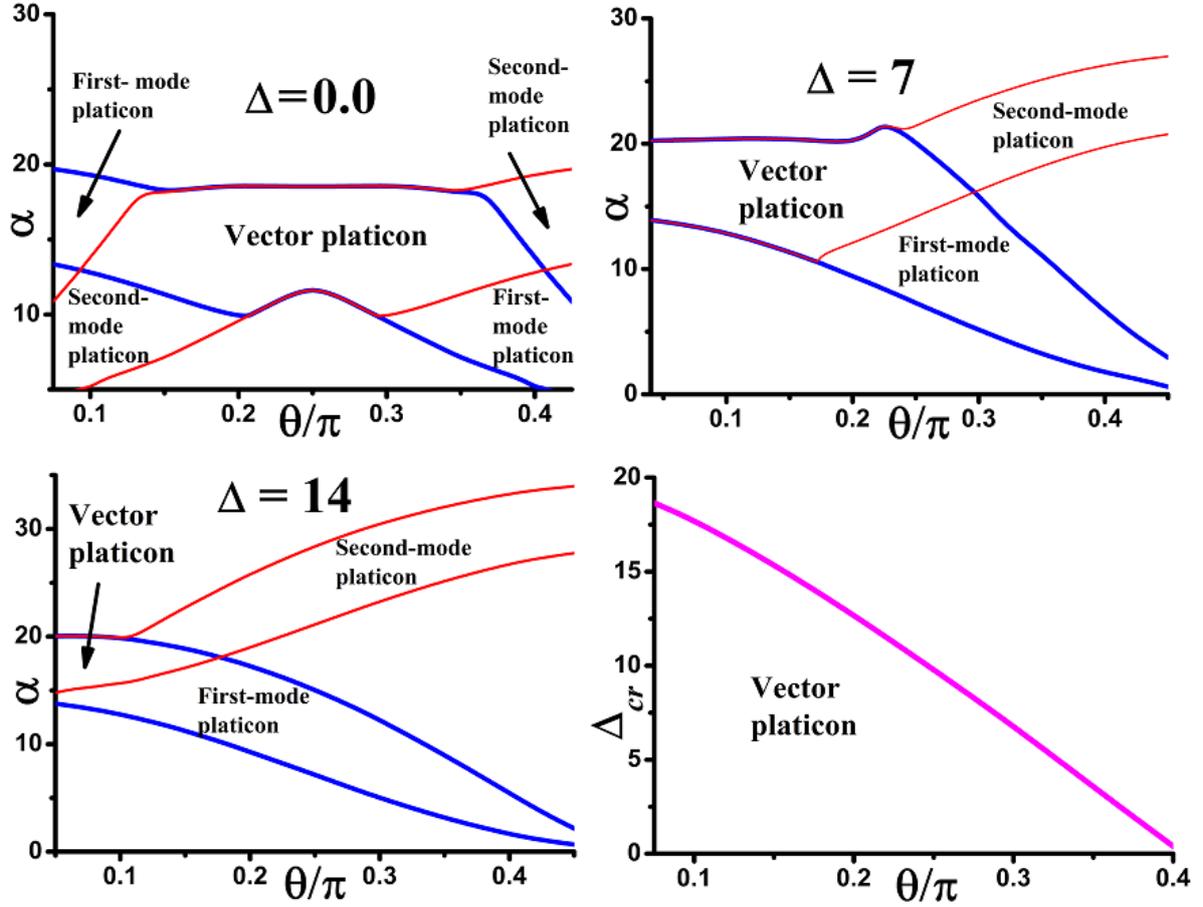

**Fig. 9.** Top panels and bottom left panel: First-mode (between blue lines), second-mode (between red lines) and vector platicon generation domains vs pump polarization angle $\theta$ for different intervals $\Delta$ between pumped modes. Bottom right: critical value of the spectral interval between pumped modes $\Delta_{cr}$ for the vector platicon generation vs pump polarization angle $\theta$. In all cases $d_{21} = d_{22} = -0.02$, $f = 6.0$, $\varepsilon = 0.4$, $v = 0.0025$, $\delta_{1,2} = 0$. All quantities are plotted in dimensionless units.

We also studied the influence of the difference of the pumped modes FSRs on the efficiency of vector platicon excitation. We considered the case when modulation frequency is equal to the first-mode FSR ($\delta_1 = 0$) and is detuned from the second-mode FSR ($\delta_2 \neq 0$). With the growth of the offset value the vector platicons generation domain becomes narrower in terms of the pump frequency detuning values (see top panels in Fig. 10). Generation of vector platicons was found to be possible for FSRs difference less than some critical value ($\delta_2 < 1$ that means that the difference should be less than $\kappa_1/2$). Note, that if $\delta_2 \neq 0$, second-mode platicon profile (blue line) becomes asymmetric (compare bottom panels in Fig. 10).

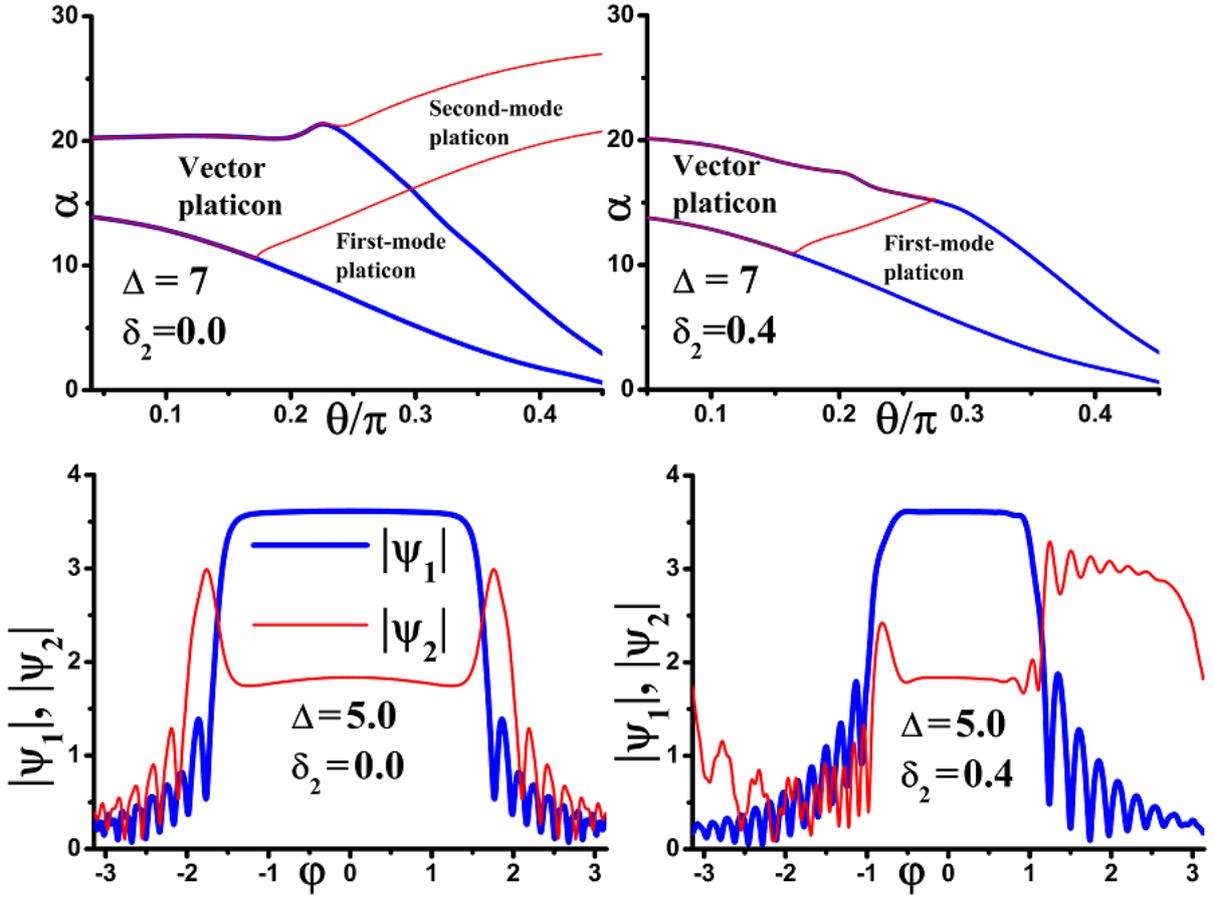

**Fig. 10.** Top panels: Vector platicon generation domains for $\delta_2 = 0$ (left panel) and $\delta_2 = 0.4$ (right panel) at $\Delta = 7.0$. Bottom panels: Vector platicon components profiles for $\delta_2 = 0$ (left panel) and $\delta_2 = 0.4$ (right panel) at $\Delta = 5.0$, $\alpha = 14.0$. In all cases $d_{21} = d_{22} = -0.02$, $f = 6.0$, $\varepsilon = 0.4$, $v = 0.0025$, $\delta_1 = 0$. All quantities are plotted in dimensionless units.

It was shown in [29] that generation of conventional platicons by modulated pump is possible if the pump amplitude is less than some critical value depending on the modulation depth. For $\varepsilon = 0.4$ platicon excitation is absent if pump amplitude $f > 8.5$. In the case of vector platicon generation one may use larger pump amplitudes since pump power is redistributed between two pumped modes. With the growth of the pump amplitude above $f = 6$ generation domain becomes smaller and if $f > 9$ it is split into two areas (see Fig. 11).

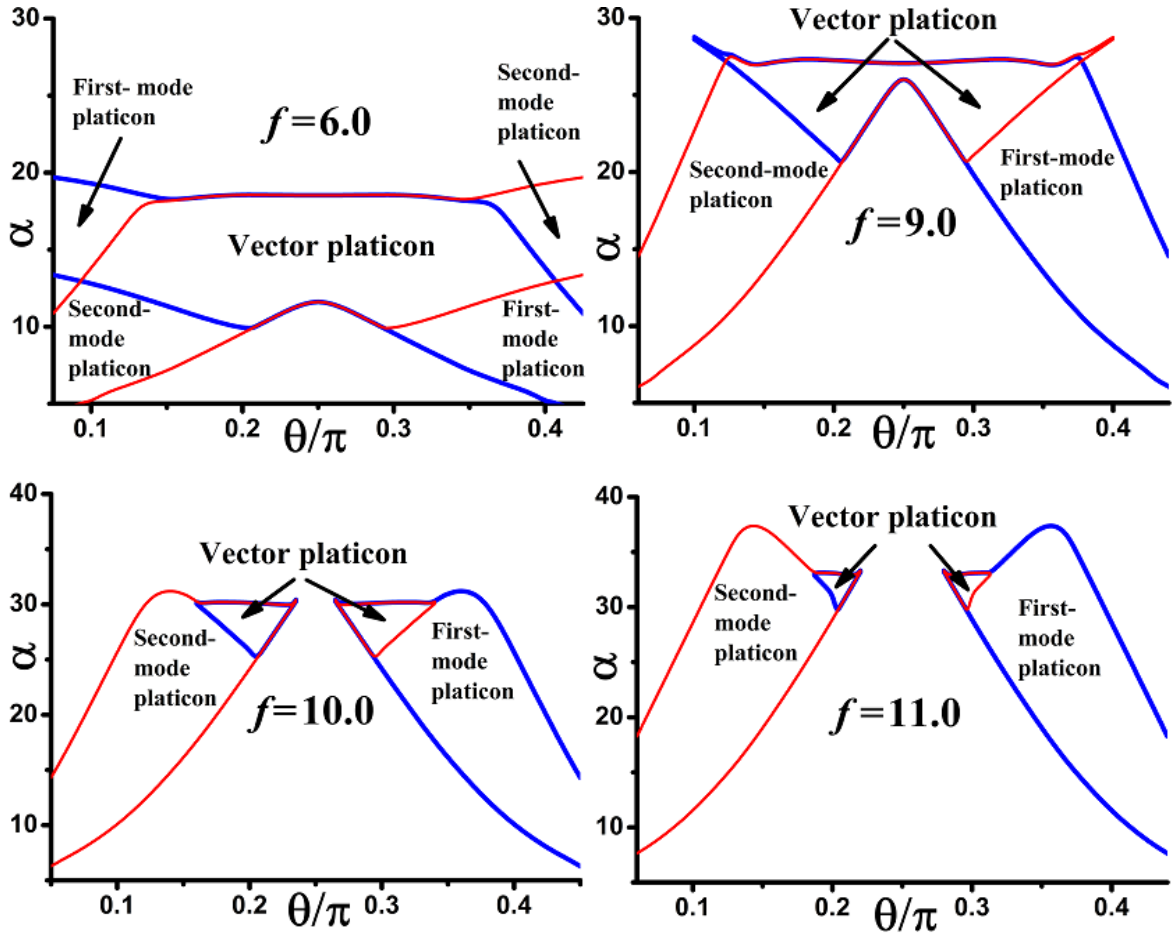

**Fig. 11.** Vector platicon generation domains for different pump amplitudes at $d_{21} = d_{22} = -0.02$, $\varepsilon = 0.4$, $v = 0.0025$, $\Delta = 0$, $\delta_{1,2} = 0$. All quantities are plotted in dimensionless units.

We found that for the generation of the vector platicons the critical pump amplitude is approximately $\sqrt{2}$ times larger than for the generation of the conventional scalar platicons. Thus, for the pump modulation depth $\varepsilon = 0.4$ the generation of vector platicons was observed for $f < 12$.

Note, that generation domains also depend on the spectral interval $\Delta$ between pumped modes and two domains may transform into one if this interval is large enough (compare Fig. 12 and bottom left panel in Fig. 11).

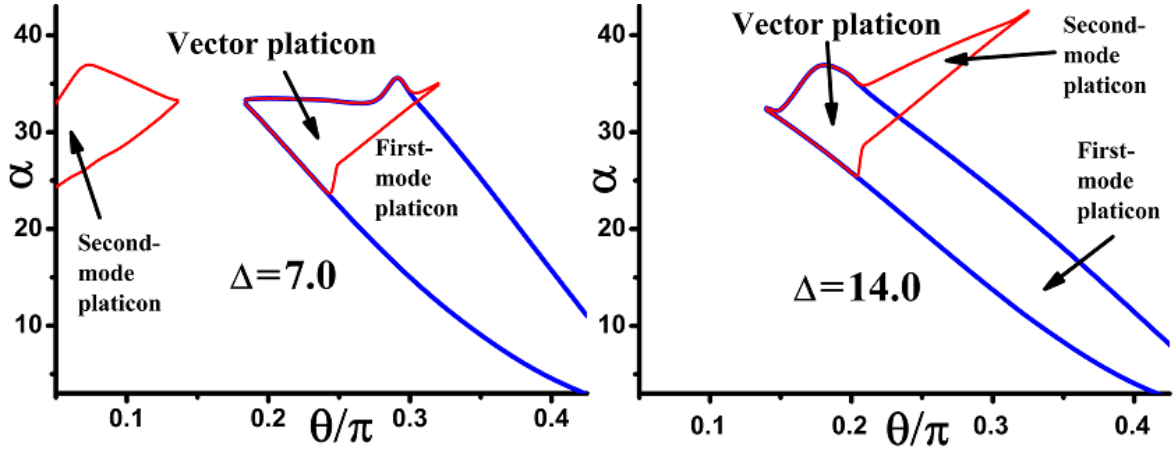

**Fig. 12.** Vector platicon generation domains for different intervals between pumped modes at $d_{21} = d_{22} = -0.02$, $f = 10$, $\varepsilon = 0.4$, $v = 0.0025$, $\delta_{1,2} = 0$. All quantities are plotted in dimensionless units.

## V. HYBRID SOLITON-PLATICON COMPLEXES

We also considered a more exotic case, when one mode family experiences anomalous GVD, but another has normal GVD. In that case one may observe simultaneous action of two effects: generation of the single-soliton or two-soliton state due to the pump modulation described in [41], and platicon generation by the modulated pump [29,30]. If the modulation depth $\varepsilon$ is large enough and the scan rate $v$ is small enough, one or two solitons are generated [41]. Generation of two-soliton state occurs due to the symmetry-breaking effect described in [42] when soliton attraction point shifts away from the minima of driving field amplitude inhomogeneities and, thus, two attraction points appear. We found out that if the mode with anomalous GVD is closer to the pump ($d_{21} > 0$, $d_{22} < 0$, $\Delta > 0$), it is possible to generate hybrid nonlinearly coupled soliton-platicon structures. Due to the nonlinear coupling with the soliton component, the platicon component profile has several narrow dips (see bottom panels in Fig. 13). Interestingly, 1-soliton state survives after platicon decay, while 2-soliton solution may annihilate upon fast drop of platicon power (compare top panels in Fig. 13).

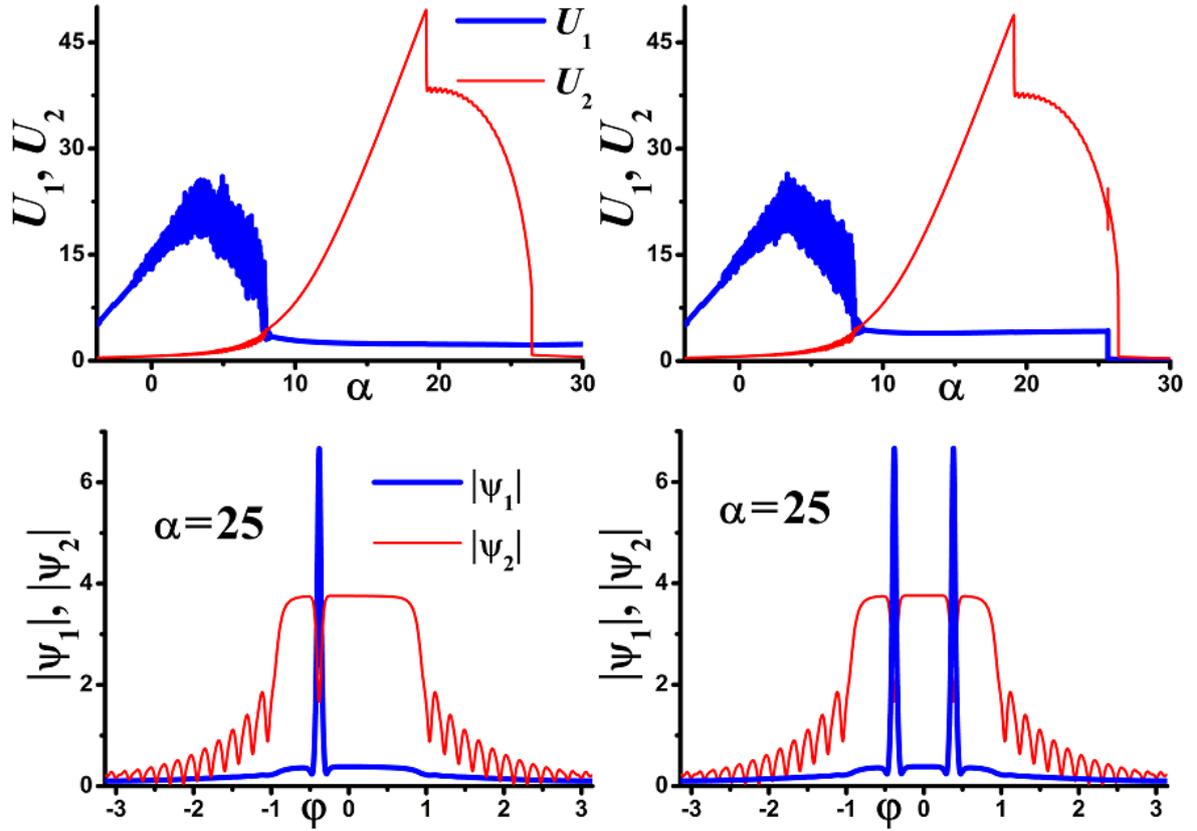

**Fig. 13.** Evolution of the intracavity powers $U_1$, $U_2$ upon pump frequency scan when the mode with anomalous GVD is pumped first (upper panels) and profiles of the generated soliton-platicon complexes with 1 and 2 solitons (bottom panels) at $d_{21} = 0.02$, $d_{22} = -0.02$, $f = 6$, $\varepsilon = 0.4$, $v = 0.0025$, $\delta_{1,2} = 0$, $\theta = \pi/4$, $\Delta = 12.0$. Bottom left profile corresponds to the top left process, bottom right profile – to the top right one. All quantities are plotted in dimensionless units.

Note, that dip position varies upon pump frequency scan moving towards the center of the platicon (see Fig. 14). Such soliton-platicon complexes were studied in [43] for the dual-pump setting. It should be noted that for the realization of such process the pump frequency scan rate $v$ should be small enough: for considered parameters $v \leq 0.005$.

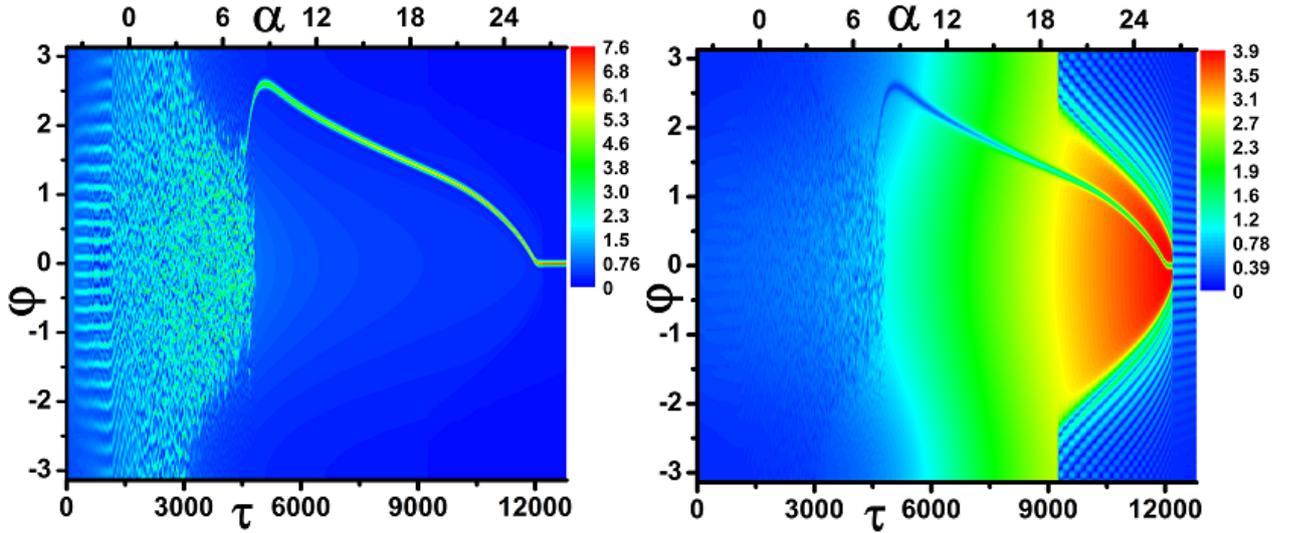

**Fig. 13.** Field distribution evolution upon pump frequency scan for the first (left panel) and second component (right panel) at $d_{21} = 0.02$, $d_{22} = -0.02$, $f = 6$, $\varepsilon = 0.4$, $v = 0.0025$, $\delta_{1,2} = 0$, $\theta = \pi/4$, $\Delta = 12.0$. All quantities are plotted in dimensionless units.

Generation of such hybrid complexes occurs for the particular range of values of the interval $\Delta$ between pumped modes: it should be large enough to provide generation of platicon at larger detuning values than generation of solitons, but also providing the overlapping of their generation ranges. The optimal range of the interval between pumped modes strongly depends on the polarization angle $\theta$ (see Fig. 15). For example, for $d_{21} = 0.02$, $d_{22} = -0.02$, $f = 6.0$, $\varepsilon = 0.4$, $v = 0.0025$ the generation was observed at $5 < \Delta < 23.5$ for $\theta = 0.3\pi$ and at $8.5 < \Delta < 13.0$ for $\theta = 0.15\pi$. For platicon generation effective the second-mode pump should be less than a critical value: for $\varepsilon = 0.4$ it should be $f \sin\theta < 8.5$. Also, to avoid the influence of the transient chaos effect described in [44] and subsequent decay of soliton at slow frequency scan, the first-mode pump amplitude also should be less than some critical value: in the considered case $f\cos\theta < 6.4$. These conditions show that generation of soliton-platicon complexes is possible for the limited range of the pump polarization angles dependent on the pump amplitude. This process was also found to be very sensitive to FSRs difference: for $\delta_1 = 0$ the generation of such structures was observed just for $\delta_2 < 0.05$.

If normal GVD mode is pumped first ($d_{21} < 0$, $d_{22} > 0$, $\Delta > 0$), generation of such structures mostly does not occur: generation of bright soliton mostly takes place after platicon decay. The excitation of hybrid complexes in that case was found for a narrow range of polarization angles close to $\theta = \pi/2$ (see Fig. 15).

It should be noted that if different signs of the GVD coefficient are realized close to the zero-dispersion point, then high-order dispersion terms should be taken into account, since they can have a strong influence on the dynamics and properties of both solitons [41-44] and platicons [49-51].

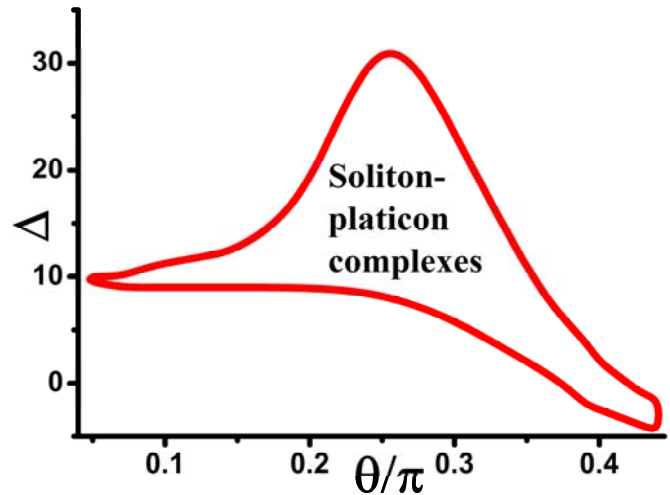

**Fig. 15.** Parameter range providing generation of soliton-platicon complexes at $d_{21} = 0.02$, $d_{22} = -0.02$, $f = 6$, $\varepsilon = 0.4$, $v = 0.0025$, $\delta_{1,2} = 0$ (the inner area between thick red curves). All quantities are plotted in dimensionless units.

## VI. CONCLUSION

We demonstrated numerically that it is possible to generate vector platicon states consisting of two nonlinearly coupled orthogonally polarized components by one amplitude-modulated pump. Generation conditions were found, and the profiles of the generated signals were analyzed. It was shown that the pump polarization provides additional degree of freedom for efficient control of the properties of the generated coherent frequency comb making them more tunable than conventional single-component platicons. It was also found that nonlinear coupling leads to the distortion of the platicon profile. We also showed the possibility of the excitation of the hybrid soliton-platicon complexes if pumped modes have alternating signs of the GVD coefficient. It was revealed that the generation of such hybrid structures, combining one platicon and one or two bright solitons is possible if anomalous GVD mode is pumped first. Reported results provide new deep insight into the complex dynamics of the nonlinear processes in high-Q microresonators and can be used to develop and create new types of microresonator-based frequency comb sources with better performance.

## ACKNOWLEDGMENT

This work was supported by the Russian Science Foundation (project 17-12-01413-П). V.E.L. and O.V.B. acknowledge personal support from the Foundation for the Advancement of Theoretical Physics and Mathematics "BASIS".